\documentclass[twocolumn,english,aps,prl,reprint,nofootinbib,superscriptaddress]{revtex4-1}
\usepackage[T1]{fontenc}
\usepackage[latin9]{inputenc}
\usepackage{textcomp}
\usepackage{amsmath}
\usepackage{amssymb}
\usepackage{graphicx}
\usepackage{esint}

\makeatletter

\newcommand{\lyxmathsym}[1]{\ifmmode\begingroup\def\b@ld{bold}
  \text{\ifx\math@version\b@ld\bfseries\fi#1}\endgroup\else#1\fi}

\@ifundefined{textcolor}{}
{%
 \definecolor{BLACK}{gray}{0}
 \definecolor{WHITE}{gray}{1}
 \definecolor{RED}{rgb}{1,0,0}
 \definecolor{GREEN}{rgb}{0,1,0}
 \definecolor{BLUE}{rgb}{0,0,1}
 \definecolor{CYAN}{cmyk}{1,0,0,0}
 \definecolor{MAGENTA}{cmyk}{0,1,0,0}
 \definecolor{YELLOW}{cmyk}{0,0,1,0}
}

\usepackage{color}
\usepackage{babel}
\usepackage{float}
\usepackage{braket}

\makeatother

\usepackage{babel}
\begin{document}

\author{K. Kolasi\'{n}ski}

\affiliation{AGH University of Science and Technology, Faculty of Physics and
Applied Computer Science,\\
 al. Mickiewicza 30, 30-059 Kraków, Poland}

\affiliation{Institut Néel, CNRS and Université Joseph Fourier, \\
 BP 166, 38042 Grenoble, France}

\author{H. Sellier}

\affiliation{Institut Néel, CNRS and Université Joseph Fourier, \\
 BP 166, 38042 Grenoble, France}

\author{B. Szafran}

\affiliation{AGH University of Science and Technology, Faculty of Physics and
Applied Computer Science,\\
 al. Mickiewicza 30, 30-059 Kraków, Poland}

\title{Spin-orbit coupling measurement by the scanning gate microscopy}
\begin{abstract}
We propose a procedure for extraction of the Fermi surface for a two-dimensional
electron gas with a strong Rashba spin-orbit coupling from conductance
microscopy. Due to the interplay between the effective spin-orbit
magnetic field and the external one within the plane of confinement,
the backscattering induced by a charged tip of an atomic force microscope
located above the sample, leads to the spin precession, and thus to
the spin mixing of the incident and reflected modes.
 This mixing leads to a characteristic angle-dependent beating pattern
visible in the conductance maps. We show that the structure of the
Fermi level, bearing signatures of the spin-orbit coupling, can be
extracted from the Fourier transform of the interference fringes in
the conductance maps as a function of the magnetic field direction.
We propose a simple analytical model which can be used to fit the
experimental data in order to obtain the spin-orbit coupling constant.


\end{abstract}
\maketitle
\textit{Introduction.} Charge carriers in semiconductors are subject
to spin-orbit (SO) interactions \cite{Manchon2015} due to electric
fields or anisotropy of the crystal lattice. The consequences of these
interactions, including spin relaxation and dephasing \cite{Ohno1999,Dyakonov1971,Kainz2004},
spin Hall effects \cite{Hirsch1999,Sinova2004,Kato2004}, topological
insulators \cite{Koenig2007}, persistent spin helix states \cite{Bernevig2006,Koralek2009,Walser2012},
Mayorana fermions \cite{Mourik2012} etc., are extensively studied
e.g.. A spin-active devices, including spin-filters based on quantum
point contacts (QPCs) \cite{Debray2009}, spin transistors \cite{Datta1990,Schliemann2003,Zutic2004,Chuang2015,Bednarek2008},
exploiting the precession of the electron spin in the SO effective
magnetic field \cite{Meier}, are well known examples. The most popular
playground for spin effects and spin-active devices is the two-dimensional
electron gas (2DEG) in III-V heterostructures, which show a strong
built-in electric fields in the confinement layer, giving rise to
the Rashba SO coupling \cite{Bychkov1984}. The knowledge of the SO
interaction strength is of fundamental importance for description
of spin devices and phenomena. The measurements of the SO coupling
constant are usually analyzed from the Shubnikov-de Haas \cite{Nitta1997,Engels1997,Lo2002,Kwon2007,Grundler2000,Kim2010,Das1989,Park2013}
oscillations, antilocalization as observed in the magnetotransport
\cite{Koga2002}, photocurrents \cite{Ganichev2004}, or precession
of optically polarized electron spins as a function of their drift
momentum \cite{Meier2007}.

The SO coupling produces a shift of the spin-up and spin-down dispersion
relations on the wave vector scale \cite{Manchon2015} that is a linear
function of the SO coupling constant. In this Letter we propose a
way to extract the structure of the dispersion relation near the Fermi
level \cite{Manchon2015} using spin-dependent scattering and the
resulting interference with the scanning gate microscopy \cite{Sellier2011,Ferry2011}
(SGM) applied to systems with QPCs \cite{Topinka2001,Schnez2011}.
In this technique, the tip acts as a floating perturbation of the
potential landscape as seen by the Fermi level electrons. As a result
the recorded SGM images contain interference fringes due to the incident
and backscattered electron waves \cite{Jura2007,Jura2009}. In presence
of an in-plane magnetic field the fringes form beating pattern due
to spin-dependence of the Fermi wavelengths \cite{Kleshchonok2015}.
In this Letter we analyze the beating patterns that appear for SO-coupled
systems. The electron -- when scattered -- experiences precession
of its spin due to rotation of the momentum-dependent effective magnetic
field \cite{Meier2007}, and the interference of the incident and
reflected electron waves potentially involves spin-mixing effects.
However, we find that in the absence of the external magnetic field
the backscattering involves a pure inversion of the effective field
with no precession effect. The latter are triggered by an external
in-plane magnetic field, and lead to an appearance of the dependence
of the beating patterns on the orientation of the magnetic field.
We demonstrate that the shape of the Fermi level structure and thus
the SO coupling constant can be traced back from the beating patterns
by Fourier transform analysis.

\begin{figure}[h]
\begin{centering}
\includegraphics[width=0.7\columnwidth]{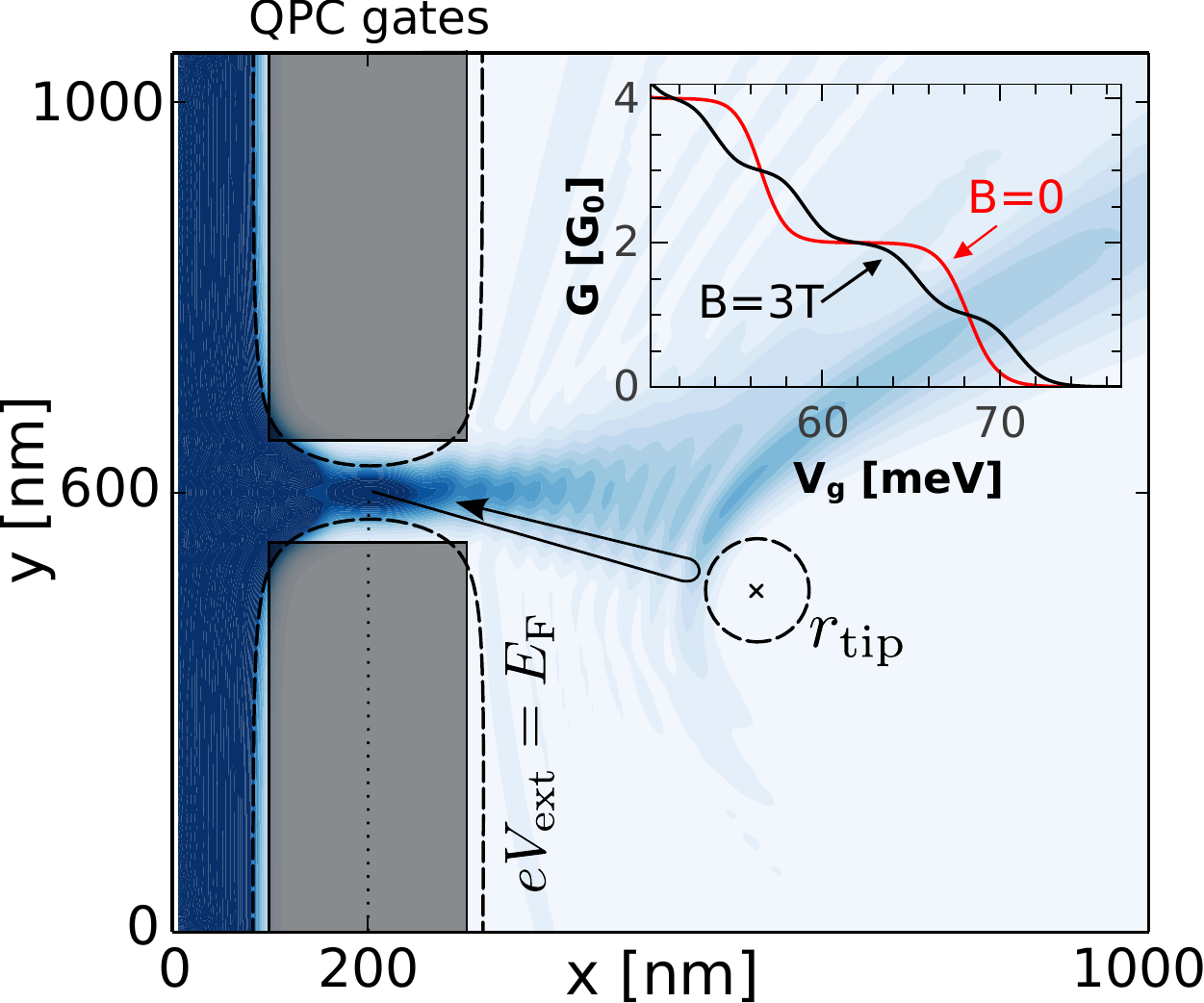} 
\par\end{centering}

\caption{\label{fig:qpcmodel}Sketch of system. The electrons come from the
reservoir on the left of the QPC. The computational box starts at
$x=0$nm. The QPC gates form a gap of size $200\mathrm{nm}\times100$nm
centered at $\left(200\mathrm{nm},600\mathrm{nm}\right)$ . The gates
are located at $50$nm above the 2DEG layer \cite{Davies1995}. Dashed
lines show the potential energy isolines for which $eV_{\mathrm{ext}}=E_{\mathrm{F}}$
in leads. The SGM tip is located at ${\bf r}_{\mathrm{tip}}=\left(x_{\mathrm{tip}},y_{\mathrm{tip}},50\mathrm{nm}\right).$
The blue map shows (the square root of) the scattering electron density
obtained for the QPC tuned to the first conductance plateau for electrons
incident from the left lead for $B=0$. The inset presents the standard
conductance quantization as a function of gate voltage $V_{\mathrm{g}}$
in case without and with in-plane magnetic field.}
\end{figure}

\textit{Theory.} We consider Fermi level transport in a 2DEG with
a local constriction QPC as depicted in Fig. \ref{fig:qpcmodel}.
The Fermi level electrons travel from the electron reservoir placed
at $x<100$ nm through a channel modeled with an infinite potential
step and an additional potential tuned by gates (gray areas of the
scheme). A negatively charged tip acts as a backscatterer to the right
of the QPC. The conductance maps as functions of the tip position
resolve the coherent interference fringes as observed in a number
of experiments \cite{Topinka2000,Topinka2001,Jura2007,Jura2009,Kozikov2015}.
The part of the system to the right of the QPC is considered open
such that electron may freely propagate without reflections. Transparent
boundary conditions for the electron flow are introduced with a method
described in Ref. \cite{Nowak2014}.

We adopt a standard two-dimensional model assuming that all the electrons
of 2DEG occupy a strongly localized lowest-energy state of the vertical
quantization. The Hamiltonian accounts for the Rashba SO interaction
and a presence of the external magnetic field applied within the plane
of confinement 
\begin{align}
H=\left[\frac{\hbar^{2}}{2m_{\mathrm{eff}}}\boldsymbol{k}^{2}+eV_{\mathrm{ext}}({\bf r})\right]\boldsymbol{I}+\frac{1}{2}g\mu_{\mathrm{B}}\boldsymbol{B}\cdot\boldsymbol{\sigma}+\boldsymbol{H}_{\mathrm{rsb}}\label{eq:mainHam}
\end{align}
 with ${\bf k}=-i\boldsymbol{\nabla}-e\boldsymbol{A}$, $\boldsymbol{B}=\left(B_{\mathrm{x}},B_{\mathrm{y}},0\right)$,
and $\boldsymbol{\sigma}$ is the vector of Pauli matrices. The external
potential $V_{\mathrm{ext}}$ is a superposition of two components:
\textit{(i)} $V_{\mathrm{QPC}}$ -- the QPC gate potential modeled
with analytical formulas for a rectangle gate adapted from Ref. \cite{Davies1995},
and \textit{(ii)} $V_{\mathrm{tip}}$ -- the electrostatic potential
created by the charged tip of the scanning probe. The tip potential
is modeled by the Lorentzian profile given by $V_{\mathrm{tip}}=d_{\mathrm{tip}}^{2}V_{\mathrm{t}}/\left[\left(x-x_{\mathrm{tip}}\right)^{2}+\left(y-y_{\mathrm{tip}}\right)^{2}+d_{\mathrm{tip}}^{2}\right],$
with effective width $d_{\mathrm{tip}}=50$nm, which is of order of
the distance between 2DEG and surface of the sample, and $V_{t}$
that depends on the voltage applied to the tip. This form of the potential
results from the screening of the tip charge by 2DEG \cite{kolasinskiDFT2013,Steinacher2015}.
The Rashba Hamiltonian $\boldsymbol{H}_{\mathrm{rsb}}=\gamma\left\{ \boldsymbol{\sigma}_{x}{k}_{y}-\boldsymbol{\sigma}_{y}{k}_{x}\right\} $
in Eq. \eqref{eq:mainHam} comes from the electrostatic confinement
of the 2DEG in the growth direction \cite{Bhandari2013}. We apply
the symmetric gauge $\boldsymbol{A}=(B_{{y}}z,\lyxmathsym{\textminus}B_{{x}}z,0)$.
By choosing the plane of the 2DEG confinement to be located at $z=0$,
we get $\boldsymbol{A}=\boldsymbol{0}$, and the magnetic field enters
the Hamiltonian only via the spin Zeeman term.

The scattering problem is solved within the finite difference approach
\cite{Kolasinski2016Lande} with spatial discretization $\Delta x=\Delta y=6$nm
using the wave function matching (WFM) method \cite{Zwierzycki2008}.
Then we calculate conductance $G$ using the Landauer approach by
evaluating $G=G_{\mathrm{0}}\sum_{\sigma}T_{\sigma}$ at the Fermi
level (with $G_{\mathrm{0}}=\frac{e^{2}}{h}$). For simplicity, we
consider the case of single mode transmitting through the QPC ($G\leq2G_{\mathrm{0}}$)
(see the inset to Fig. 1). We set $E_{\mathrm{F}}=20$meV (for $\gamma=0$
the Fermi wavelength is $\lambda_{\mathrm{F}}=40$nm), and the tip
potential $V_{\mathrm{t}}=40$meV for which a strict depletion of
the electron density below the tip is obtained (see dashed circle
in Fig. \ref{fig:qpcmodel}). Landé factor is assumed to be $g=9$
and effective mass $m_{\mathrm{eff}}=0.0465m_{0}$ as for InGaAs.

\begin{figure}[h]
\begin{centering}
\includegraphics[width=1\columnwidth]{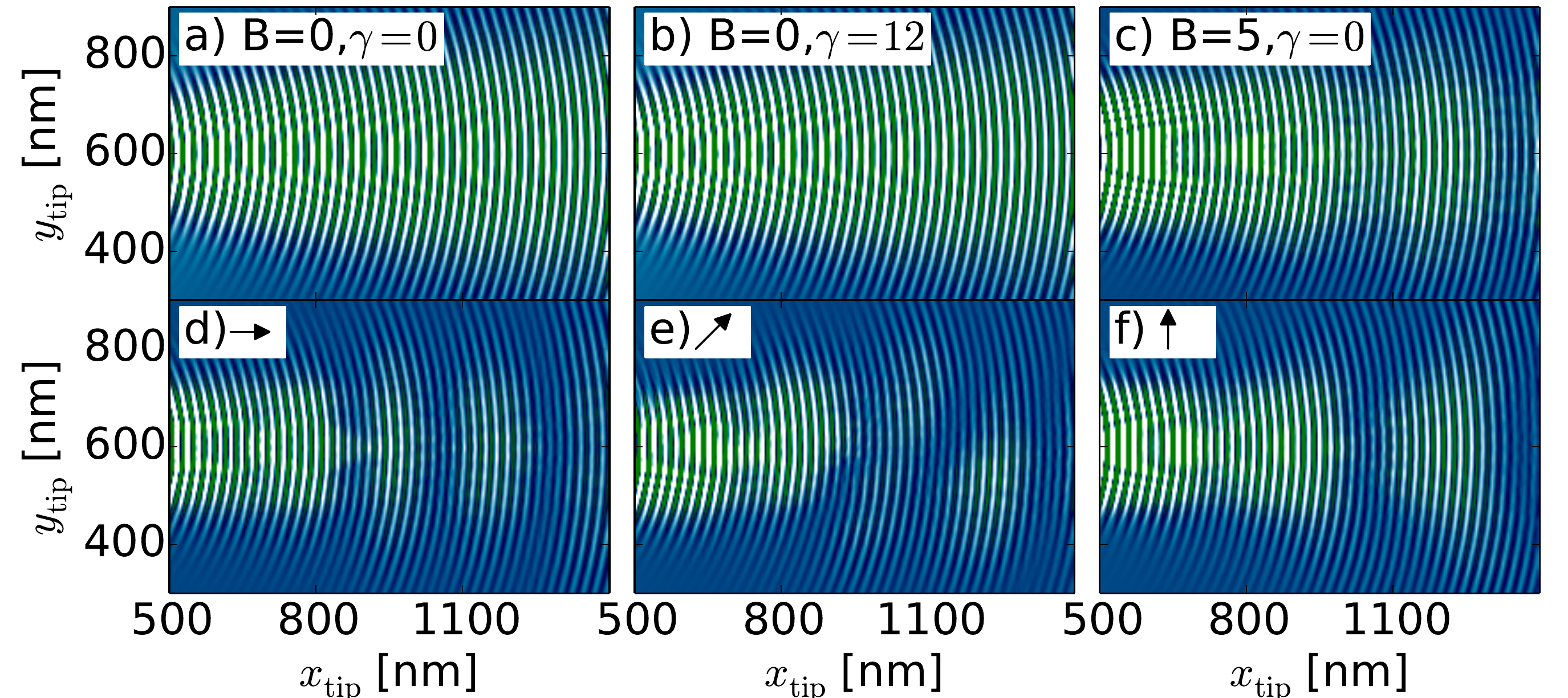} 
\par\end{centering}

\caption{\label{fig:sgms} Derivatives of simulated SGM images ($dG/dx_{\mathrm{tip}}$)
obtained for QPC tuned to the first plateau in arb. units. $dG/dx_{\mathrm{tip}}$
map obtained in absence of external magnetic field and SO interaction
(a), with SO coupling ($\gamma=12$meVnm) at $B=0$ (b), for in-plane
magnetic field $B=5$T and without SO interaction (c). (d-f) $dG/dx_{\mathrm{tip}}$
images obtained for in-plane magnetic field $B=5$T and $\gamma=12$meVnm.
The arrows show the in-plane direction of the $\boldsymbol{B}$ vector.}
\end{figure}

\textit{Results and Discussion.} Figs. \ref{fig:sgms}(a-f) show spatial
derivatives of SGM images $dG/dx_{\mathrm{tip}}$ obtained from the
solution of the quantum scattering problem for QPC in Fig. \ref{fig:qpcmodel}
tuned to the first QPC conductance plateau. 
For $B=0$ and $\gamma=0$ {[}Fig. 2(a){]} a pronounced interference
pattern of the incident and backscattered wave is observed \cite{Topinka2001,Schnez2011,Jura2007,Jura2009}
with the period of $\lambda_{F}/2$ for both $\gamma=0$ {[}Fig. 2(a){]}
and $\gamma\neq0$ {[}Fig. 2(b){]}. A beating pattern \cite{Kleshchonok2015}
appears at non-zero $B$ {[}Fig. \ref{fig:sgms}(c){]}, which depends
on the orientation of the in-plane field for $\gamma\neq0$ (Figs.
\ref{fig:sgms} (d-f)).

\begin{figure}[h]
\begin{centering}
\includegraphics[width=0.45\columnwidth]{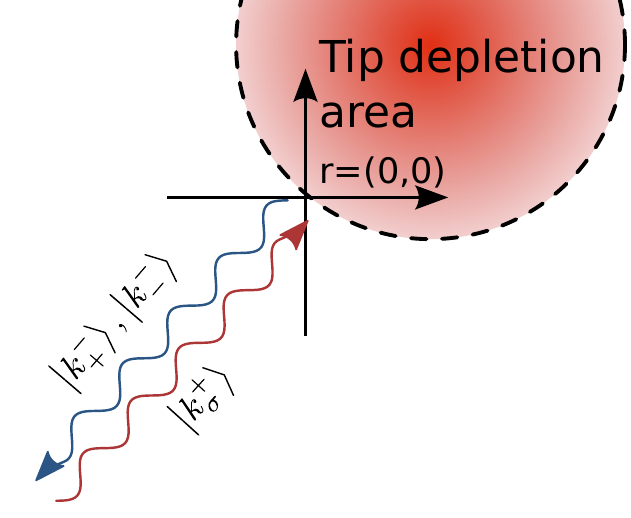} 
\par\end{centering}

\caption{\label{fig:scattering}Sketch of considered scattering process. The
electron wave leaves QPC in one of two spin states, propagates to
the right and is backscattered at position $r=(0,0)$ by the potential
barrier induced by the SGM tip. Here we assume a hard wall potential
profile (i.e. $V_{\mathrm{tip}}=+\infty$ inside the circle).}
\end{figure}

In order to explain the results of Fig. 2 we consider a simple model
for SGM images in presence of in-plane magnetic field and SO interaction.
The electron wave which leaves the QPC \cite{Kolasinski2014Slit,Kolasinski2015,Jalabert2010,Khatua2014}
is approximated by a plain wave $e^{ikr}$ (an inverse of the square
root of the distance from the QPC is neglected as slowly varying).
The schematics of the considered scattering process is presented in
Fig. \ref{fig:scattering}. The electron wave which leaves the QPC
(not shown in the diagram) propagates through the device until it
is backscattered by the potential barrier created by the SGM tip with
probability 1. We fix the origin at the scattering point. 
 For a given incoming spin state $\left|k_{\sigma}^{+}\right\rangle $
the scattering wave function can be expanded in terms of the possible
scattering modes 
\begin{equation}
\left|\Psi_{\sigma}\right\rangle =e^{ik_{\sigma}^{+}r}\left|k_{\sigma}^{+}\right\rangle +\Sigma_{\sigma'}a_{\sigma\sigma'}e^{-ik_{\sigma'}^{-}r}\left|k_{\sigma'}^{-}\right\rangle ,\label{eq:scatteq}
\end{equation}
 
where $k_{\sigma}^{\pm}r=\left|\boldsymbol{k}_{\sigma}^{\pm}\cdot\boldsymbol{r}\right|$
and $k_{\sigma}^{\pm}$ denotes the absolute value of the wave vector
of an electron in spin state $\sigma$. The sign in the superscript
indicates the electron incoming from left $+$ or backscattered by
the tip $-$. The values of the scattering amplitudes $a_{\sigma\sigma'}$
depend on a specific situation.
For SO coupling and magnetic field simultaneously present, the Hamiltonian
for a free electron can be written {\small { 
\begin{equation}
\boldsymbol{H}=\left[\begin{array}{cc}
\boldsymbol{E}_{\mathrm{kin}} & \gamma\left(\boldsymbol{k}_{\mathrm{y}}+i\boldsymbol{k}_{\mathrm{x}}\right)+\alpha_{\mathrm{x}}-i\alpha_{\mathrm{y}}\\
\gamma\left(\boldsymbol{k}_{\mathrm{y}}-i\boldsymbol{k}_{\mathrm{x}}\right)+\alpha_{\mathrm{x}}+i\alpha_{\mathrm{y}} & \boldsymbol{E}_{\mathrm{kin}}
\end{array}\right],\label{eq:hamm}
\end{equation}
}}where $\alpha_{x/y}=\frac{1}{2}g\mu_{B}B_{x/y}$ and $\boldsymbol{E}_{\mathrm{kin}}=\frac{\hbar^{2}\boldsymbol{k}^{2}}{2m_{\mathrm{eff}}}$.
Plain wave solution for the Schrödinger equation 
gives two eigenvalues 
\begin{equation}
E_{\sigma}=\frac{\hbar^{2}\boldsymbol{k}^{2}}{2m_{\mathrm{eff}}}+\sigma\left|\boldsymbol{p}\right|,\label{eq:reldysp}
\end{equation}
where $\boldsymbol{p}=\left(\gamma k_{\mathrm{y}}+\alpha_{\mathrm{x}},-\gamma k_{\mathrm{x}}+\alpha_{\mathrm{y}}\right)$,
with $\sigma=\{+,-\}$ and 
\begin{equation}
\left|k_{\sigma}^{\pm}\right\rangle =\frac{1}{\sqrt{2}}\left(\begin{array}{c}
1\\
\sigma\frac{p_{\mathrm{x}}^{\pm}+ip_{\mathrm{y}}^{\pm}}{p^{\pm}}
\end{array}\right),\label{eq:modes}
\end{equation}
where $p^{\pm}$ denotes the value of $\boldsymbol{p}$ vector, are
eigenvectors for incoming $+$ and outgoing directions $-$ of an
electron. 
 Due to the assumed infinite potential generated by the SGM tip, the
scattering wave function in Eq. \eqref{eq:scatteq} has to vanish
at $r=0$ (see Fig. \ref{fig:scattering}) 
\[
\Psi_{\sigma}\left(r=0\right)=\left|k_{\sigma}^{+}\right\rangle +\Sigma_{\sigma'}a_{\sigma\sigma'}\left|k_{\sigma'}^{-}\right\rangle =0.
\]
 By substituting Eq. \eqref{eq:modes} to this equation one evaluates
the scattering amplitudes $a_{\sigma\sigma'}$. 

\begin{figure}[h]
\begin{centering}
\includegraphics[width=1\columnwidth]{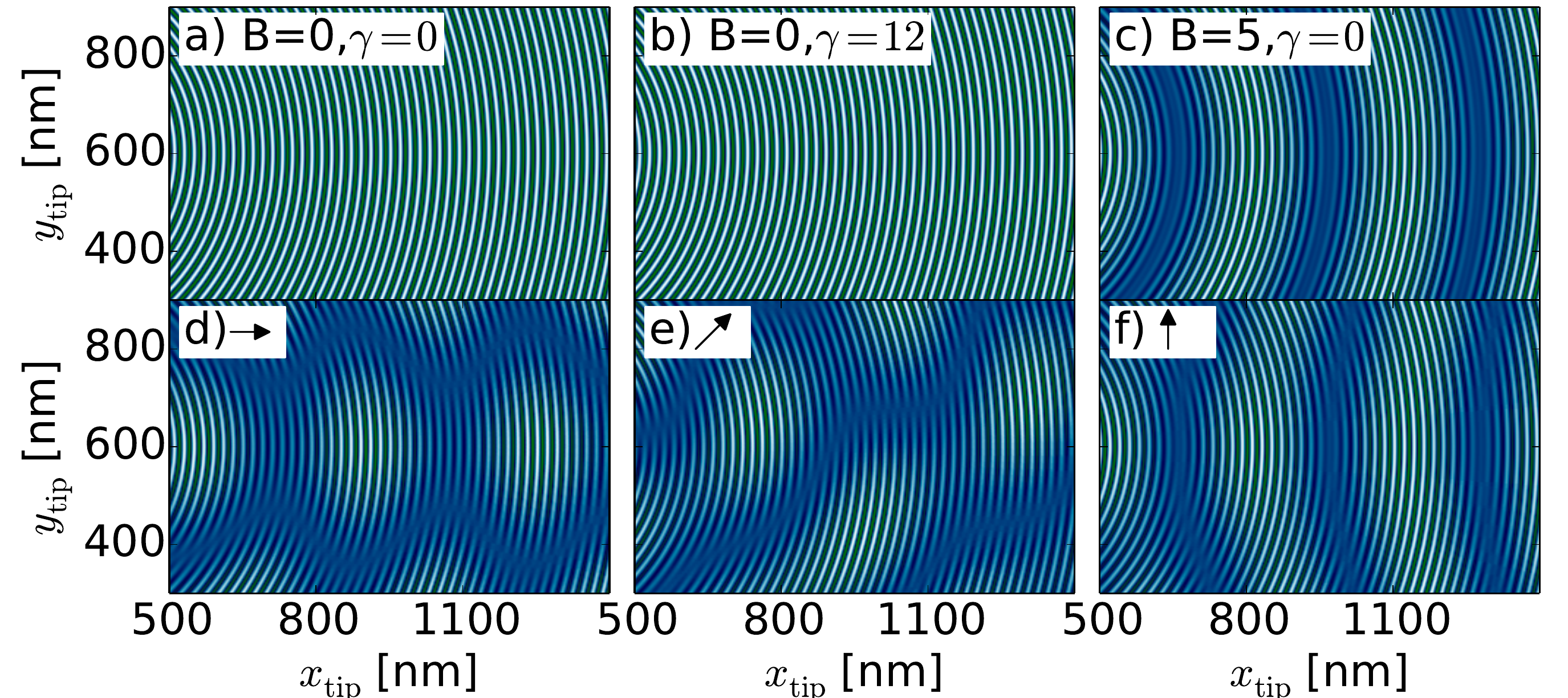} 
\par\end{centering}

\caption{\label{fig:sgmm}Same as on Fig. \ref{fig:sgms} but calculated from
simple model discussed in this paper.}
\end{figure}

For the simplest case when SOI and magnetic field are not present
in the Hamiltonian \eqref{eq:hamm} the propagating modes in Eq. \eqref{eq:modes}
reduce to 
\[
\left|k_{+}^{\pm}\right\rangle =\left|k_{+}\right\rangle =\left(\begin{array}{c}
1\\
0
\end{array}\right),\,\,\left|k_{-}^{\pm}\right\rangle =\left|k_{-}\right\rangle =\left(\begin{array}{c}
0\\
1
\end{array}\right),
\]
 with $k_{\sigma}^{\pm}=k$ and scattering amplitudes $a_{\sigma\sigma'}=-\delta_{\sigma\sigma'}$,
from which one finds that reflection does not change the spin orientation.
The scattering wave function from Eq. \eqref{eq:scatteq} is then
\[
\left|\Psi_{\sigma}\right\rangle =e^{ikr}\left|k_{\sigma}^{+}\right\rangle -e^{-ikr}\left|k_{\sigma}^{-}\right\rangle =\left(e^{ikr}-e^{-ikr}\right)\left|k_{\sigma}\right\rangle ,
\]
 and the scattering density is given by $\rho_{\sigma}=\ensuremath{\braket{\Psi_{\sigma}|\Psi_{\sigma}}}\propto\cos\left(2kr\right),$
and the variation of the $G$ map follows the pattern of the density
\cite{Kolasinski2014Slit}, The SGM conductance pattern can be approximated
by 
$G\left(r_{\mathrm{tip}}\right)\propto\cos\left(2kr\right)$. The
SGM image obtained with this model is presented in Fig. \ref{fig:sgmm}(a)
and is consistent with the simulated image obtained in Fig. \ref{fig:sgms}.

For $B=0$ and $\gamma\neq0$ one may easily check that the propagating
modes {[}Eq. (5){]} still satisfy orthogonality relations $\braket{k_{\sigma'}^{+}|k_{\sigma}^{-}}=\delta_{\sigma\sigma'}$
and $\left|k_{\sigma}^{+}\right\rangle =\left|k_{\sigma}^{-}\right\rangle $,
which leads to the spin conserving reflection $a_{\sigma\sigma'}=-\delta_{\sigma\sigma'}$.
However in this case $k_{\sigma}^{+}\neq k_{\sigma}^{-}$ and the
scattering wave function is given by $\left|\Psi_{\sigma}\right\rangle =\left(e^{ik_{\sigma}^{+}r}-e^{-ik_{\sigma}^{-}r}\right)\left|k_{\sigma}^{+}\right\rangle $.
The electron density is then proportional to $\rho_{\sigma}\propto\cos\left(\left[k_{\sigma}^{+}+k_{\sigma}^{-}\right]r\right)$.
However, using the fact that $k_{\sigma}^{\pm}=k^{\pm}+\sigma\frac{\gamma m_{\mathrm{eff}}}{\hbar^{2}}$
\cite{Bercioux2015} we get the same expression as for $\gamma=0$
i.e. $\rho_{\sigma}\propto\cos\left(2kr\right)$, which does not depend
on electron spin. Hence the SO effect vanishes for the backscattering
process which leads to the same SGM image {[}Fig. \ref{fig:sgmm}(b){]}
as in case of $\gamma=0$ {[}Fig. \ref{fig:sgmm}(a){]}. 


The third possible configuration of parameters i.e. $\gamma=0$ and
$B\neq0$ was recently discussed in Ref. \cite{Kleshchonok2015}.
In this case the same orthogonality relation is still satisfied $\ensuremath{\braket{k_{\sigma'}^{+}|k_{\sigma}^{-}}}=\delta_{\sigma\sigma'}$,
and $a_{\sigma\sigma'}=-\delta_{\sigma\sigma'}$. However, the resulting
electron density is now proportional to $\rho_{\sigma}\propto\cos\left(2k_{\sigma}r\right)$
and depends on the spin via the Zeeman term in Eq. (4) inducing shifts
of $k_{\sigma}$. The approximated SGM map $G=G_{\mathrm{0}}\sum_{\sigma}T_{\sigma}\cos\left(2k_{\sigma}r\right)$
 gives a signal being a superposition of two frequencies $\omega_{\sigma}=2k_{\sigma}$
resulting in the beating pattern visible in Fig. \ref{fig:sgmm}(c).
The present reasoning explains the findings of Ref. \cite{Kleshchonok2015}.


In a general case of $B\neq0$ and $\gamma\neq0$ the eigenvalues
\eqref{eq:reldysp} depend on both the direction of magnetic field
and the propagation vector, thus the spin will not be conserved anymore
during the backscattering process, since the orthogonality relations
between the incident and backscattered modes no longer hold $\braket{k_{\sigma'}^{+}|k_{\sigma}^{-}}\neq\delta_{\sigma\sigma'}$,
and $a_{\sigma\sigma'}\neq-\delta_{\sigma\sigma'}${.}
 The resulting electron density will be then a composition of four
different possible superposition of the Fermi wave vectors $k_{i}=\left\{ k_{+}^{+}+k_{+}^{-},k_{+}^{+}+k_{-}^{-},k_{-}^{+}+k_{+}^{-},k_{-}^{+}+k_{-}^{-}\right\} $.
The SGM images obtained for this general case for three different
orientation of magnetic field $\alpha=\left\{ 0^{\circ},45^{\circ},90^{\circ}\right\} $
are depicted in Figs. \ref{fig:sgmm}(d-f). Although, the images differ
somewhat from Fig. 4 (d-f), still both the model and the full simulation
allow for extraction of the wave vectors and their dependence on the
orientation of the magnetic field in the Fourier analysis (see below).

\begin{figure}[h]
\begin{centering}
\includegraphics[width=1\columnwidth]{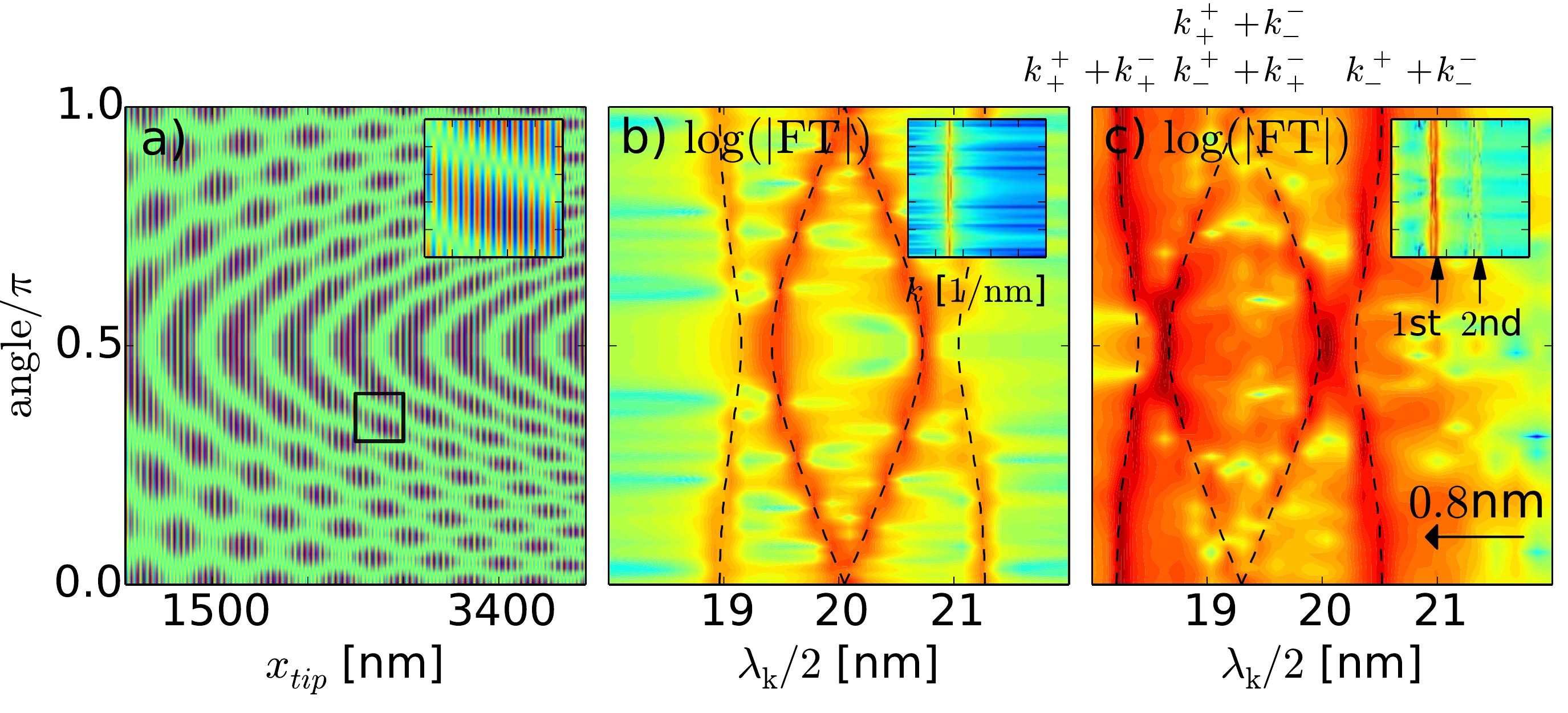} 
\par\end{centering}

\caption{\label{fig:fftAngle}a) Conductance as a function of the tip position
moving along the $x$ axis (with $y_{\mathrm{tip}}=600$nm) and the
angle that the magnetic field vector forms with the $x$ axis. The
inset shows a zoomed part in he area denoted by the black square.
The simulation was performed for 5T and $\gamma=12$meVnm with the
simple analytical model. (b) Fourier transform (FT) of (a) remapped
from $k$ space to $\lambda=2\pi/k$. Dashed lines were calculated
from the dispersion relation defined by Eq. \eqref{eq:reldysp} as
$\lambda_{i}=2\pi/k_{i}$ with $k_{i}=\left\{ k_{+}^{+}+k_{+}^{-},k_{+}^{+}+k_{-}^{-},k_{-}^{+}+k_{+}^{-},k_{-}^{+}+k_{-}^{-}\right\} $.
The inset shows the same image but in the $k$ space for a large range
of wave vectors values. c) Same as (b) only for the full numerical
simulation taken at $G=G_{\mathrm{0}}$. The finite size of the SGM
tip potential leads to shift of all lines towards higher frequencies.
Quantum mechanical simulation reveals also the higher harmonics in
the inset denoted by 1st and 2nd arrows. }
\end{figure}

The form of Eq. \eqref{eq:hamm} indicates that rotation of a SGM
tip position along the arc centered at the QPC entrance is equivalent
to a rotation of the in-plane magnetic field (in an opposite direction)
for a fixed tip position. For a practical implementation of an experiment
it should be more efficient to perform a SGM scan along a straight
line, where the longest electron branch \cite{Topinka2000,Kozikov2015}
is present and rotate the magnetic field instead (see Fig. \ref{fig:fftAngle}(a)).



\begin{figure}[h]
\begin{centering}
\includegraphics[width=1\columnwidth]{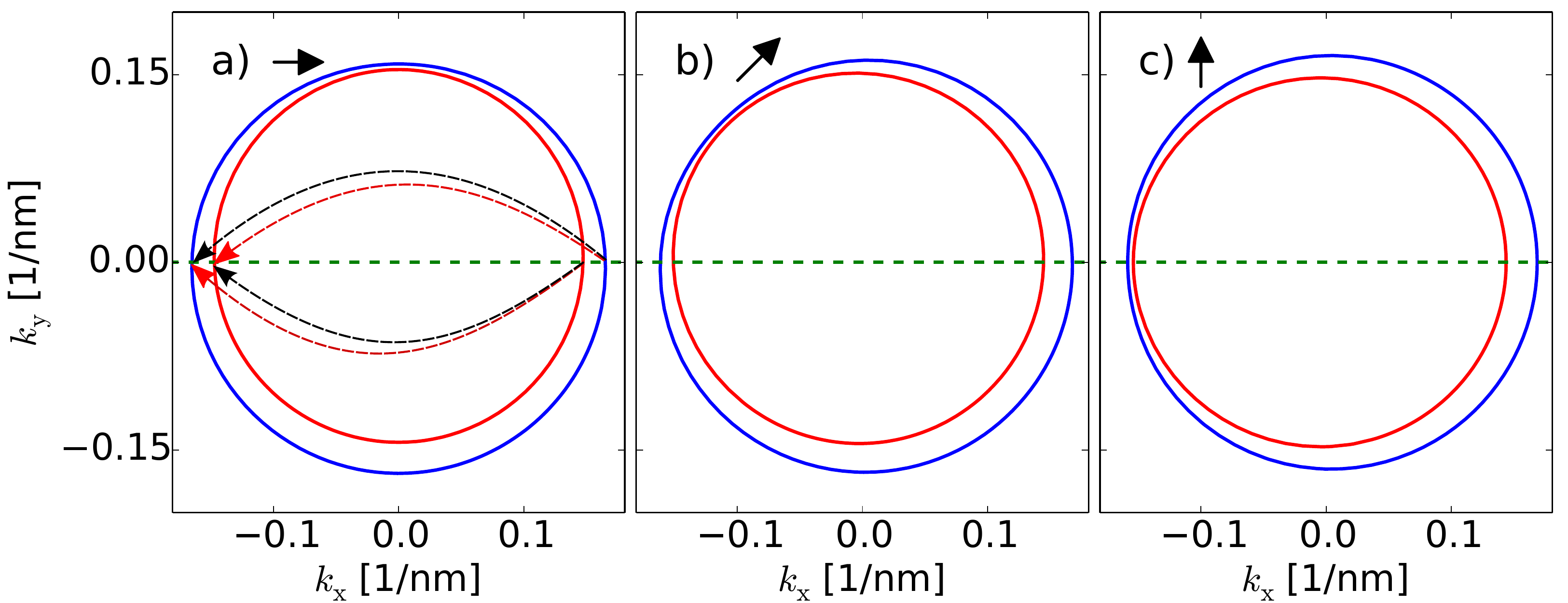} 
\par\end{centering}

\caption{\label{fig:circles}(a-c) Fermi level surface calculated from Eq.
\eqref{eq:reldysp} obtained for three directions of magnetic field
$\phi=\left\{ 0^{\circ},45^{\circ},90^{\circ}\right\} $ (for $B=5T$)
denoted by arrows. Green dashed lines show the direction of scattering
process i.e. $\boldsymbol{k}=\left(k_{\mathrm{x}},0\right)$. Dashed
arrows on (a) represent four different possible backscattering processes.
However, due to the symmetry of the scattering process two of them
lead to the same frequency in FT image, hence three lines are observed
in Fig. 5(b-c) for $\phi=0$ and $\pi$. This is no more valid for
(b) and (c), which imply four different lines in Fig. 5. }
\end{figure}

In Fig. \ref{fig:fftAngle}(b-c) we present the Fourier transform
(FT) of the conductance signal calculated from the $dG/dx_{\mathrm{tip}}$
map for the tip moving along the QPC axis, as a function of the magnetic
field direction \textbf{B} for $B=5$ T. The results are plotted on
the wavelength scale calculated as $\lambda_{F}=2\pi/k_{i}$. The
dashed lines in Figs. \ref{fig:fftAngle}(b-c) were plotted for backscattering
processes that are explained in Fig. \ref{fig:circles}(a) and calculated
numerically from the condition $E_{F}=E_{\sigma}$ with the latter
given by Eq. (4). Note, that due to the smooth and extended shape
of the the tip potential in the full simulation the resonance lines
in Fig. 5(c) are slightly shifted to the left by 0.8nm (in comparison
to model Fig. 5(b)). We accordingly shifted the dashed lines in Fig. 5(c) to coincide
with the FT image. In the inset in Fig. \ref{fig:fftAngle}(c) one
observes also higher harmonics, which result from the possible multiple
reflections between the tip and QPC (not present in the model, see
inset in Fig. 5(b)). 

The backscattering taken along the axis of the QPC involves $k_{y}=0$
and we find in general four various values of $k_{i}$ visible as
four lines in FT images. However, when $B_{y}=0$ , Eq. (4) reduces
to 
\begin{equation}
E_{\sigma}=\frac{\hbar^{2}k_{\sigma}^{2}}{2m}+\sigma\sqrt{\alpha_{\mathrm{x}}^{2}+\left(\gamma k_{\sigma}\right)^{2}},\label{eq:relBx}
\end{equation}
which is symmetric with respect to electron reflection $E_{\sigma}\left(k_{\mathrm{x}}\right)=E_{\sigma}\left(-k_{\mathrm{x}}\right)$,
which implies the symmetry o scattering process that $k_{\sigma}^{+}=k_{\sigma}^{-}\equiv k_{\sigma}$(see
Fig. 6(a)), and thus reducing the number of resonance lines in FT
image to two. For other cases presented in Figs. 6(b-c) this symmetry
is not satisfied and all four frequencies are visible.

\textit{Summary.} In summary, we have shown that SGM imaging can be
used to extract the Fermi surface properties by Fourier analysis of
the beatings due to the SO interaction and an in-plane magnetic field.
The analysis allows for deduction of the Rashba constant from the
real space measurement of conductance as a function of the tip position
involving spin-scattering in a crossed external and built-in magnetic
fields.


\textit{Acknowledgments} This work was supported by National Science
Centre according to decision DEC-2015/17/N/ST3/02266, and by PL-Grid
Infrastructure. The first author is supported by the scholarship of
Krakow Smoluchowski Scientific Consortium from the funding for National
Leading Reserch Centre by Ministry of Science and Higher Education
(Poland) and by the Etiuda stipend of the National Science Centre
(NCN) according to decision DEC-2015/16/T/ST3/00310.

 \bibliographystyle{apsrev4-1-nourl}
\bibliography{referencje}

\begin{thebibliography}{10}%
\makeatletter
\providecommand \@ifxundefined [1]{%
 \ifx #1\undefined \expandafter \@firstoftwo
 \else \expandafter \@secondoftwo
\fi
}%
\providecommand \@ifnum [1]{%
 \ifnum #1\expandafter \@firstoftwo
 \else \expandafter \@secondoftwo
\fi
}%
\providecommand \enquote [1]{``#1''}%
\providecommand \bibnamefont  [1]{#1}%
\providecommand \bibfnamefont [1]{#1}%
\providecommand \citenamefont [1]{#1}%
\providecommand\href[0]{\@sanitize\@href}%
\providecommand\@href[1]{\endgroup\@@startlink{#1}\endgroup\@@href}%
\providecommand\@@href[1]{#1\@@endlink}%
\providecommand \@sanitize [0]{\begingroup\catcode`\&12\catcode`\#12\relax}%
\@ifxundefined \pdfoutput {\@firstoftwo}{%
 \@ifnum{\z@=\pdfoutput}{\@firstoftwo}{\@secondoftwo}%
}{%
 \providecommand\@@startlink[1]{\leavevmode\special{html:<a href="#1">}}%
 \providecommand\@@endlink[0]{\special{html:</a>}}%
}{%
 \providecommand\@@startlink[1]{%
  \leavevmode
  \pdfstartlink
   attr{/Border[0 0 1 ]/H/I/C[0 1 1]}%
   user{/Subtype/Link/A<</Type/Action/S/URI/URI(#1)>>}%
  \relax
 }%
 \providecommand\@@endlink[0]{\pdfendlink}%
}%
\providecommand \url  [0]{\begingroup\@sanitize \@url }%
\providecommand \@url [1]{\endgroup\@href {#1}{\urlprefix}}%
\providecommand \urlprefix [0]{URL }%
\providecommand \Eprint[0]{\href }%
\@ifxundefined \urlstyle {%
  \providecommand \doi [1]{doi:\discretionary{}{}{}#1}%
}{%
  \providecommand \doi [0]{doi:\discretionary{}{}{}\begingroup
  \urlstyle{rm}\Url }%
}%
\providecommand \doibase [0]{http://dx.doi.org/}%
\providecommand \Doi[1]{\href{\doibase#1}}%
\providecommand \bibAnnote [3]{%
  \BibitemShut{#1}%
  \begin{quotation}\noindent
    \textsc{Key:}\ #2\\\textsc{Annotation:}\ #3%
  \end{quotation}%
}%
\providecommand \bibAnnoteFile [2]{%
  \IfFileExists{#2}{\bibAnnote {#1} {#2} {\input{#2}}}{}%
}%
\providecommand \typeout [0]{\immediate \write \m@ne }%
\providecommand \selectlanguage [0]{\@gobble}%
\providecommand \bibinfo [0]{\@secondoftwo}%
\providecommand \bibfield [0]{\@secondoftwo}%
\providecommand \translation [1]{[#1]}%
\providecommand \BibitemOpen[0]{}%
\providecommand \bibitemStop [0]{}%
\providecommand \bibitemNoStop [0]{.\EOS\space}%
\providecommand \EOS [0]{\spacefactor3000\relax}%
\providecommand \BibitemShut [1]{\csname bibitem#1\endcsname}%
\bibitem{Manchon2015}%
  \BibitemOpen
  \bibfield{author}{%
  \bibinfo {author} {\bibfnamefont{A.}~\bibnamefont{Manchon}}, \bibinfo
  {author} {\bibfnamefont{H.~C.}\ \bibnamefont{Koo}}, \bibinfo {author}
  {\bibfnamefont{J.}~\bibnamefont{Nitta}}, \bibinfo {author}
  {\bibfnamefont{S.~M.}\ \bibnamefont{Frolov}},\ and\ \bibinfo {author}
  {\bibfnamefont{D.~R.}\ \bibnamefont{A.}},\ }%
  \bibfield{journal}{%
  \bibinfo {journal} {Nat. Mater.}\ }%
  \textbf{\bibinfo {volume} {14}},\ \bibinfo {pages} {871} (\bibinfo {year}
  {2015})%
  \bibAnnoteFile{NoStop}{Manchon2015}%
\bibitem{Ohno1999}%
  \BibitemOpen
  \bibfield{author}{%
  \bibinfo {author} {\bibfnamefont{Y.}~\bibnamefont{Ohno}}, \bibinfo {author}
  {\bibfnamefont{R.}~\bibnamefont{Terauchi}}, \bibinfo {author}
  {\bibfnamefont{T.}~\bibnamefont{Adachi}}, \bibinfo {author}
  {\bibfnamefont{F.}~\bibnamefont{Matsukura}},\ and\ \bibinfo {author}
  {\bibfnamefont{H.}~\bibnamefont{Ohno}},\ }%
  \bibfield{journal}{%
  \Doi{10.1103/PhysRevLett.83.4196}{\bibinfo {journal} {Phys. Rev. Lett.}}\ }%
  \textbf{\bibinfo {volume} {83}},\ \bibinfo {pages} {4196} (\bibinfo {year}
  {1999})%
  \bibAnnoteFile{NoStop}{Ohno1999}%
\bibitem{Dyakonov1971}%
  \BibitemOpen
  \bibfield{author}{%
  \bibinfo {author} {\bibfnamefont{M.~I.}\ \bibnamefont{D'yakonov}}\ and\
  \bibinfo {author} {\bibfnamefont{V.~I.}\ \bibnamefont{Perel}},\ }%
  \bibfield{journal}{%
  \bibinfo {journal} {JETP Lett.}\ }%
  \textbf{\bibinfo {volume} {13}},\ \bibinfo {pages} {467} (\bibinfo {year}
  {1971})%
  \bibAnnoteFile{NoStop}{Dyakonov1971}%
\bibitem{Kainz2004}%
  \BibitemOpen
  \bibfield{author}{%
  \bibinfo {author} {\bibfnamefont{J.}~\bibnamefont{Kainz}}, \bibinfo {author}
  {\bibfnamefont{U.}~\bibnamefont{R\"ossler}},\ and\ \bibinfo {author}
  {\bibfnamefont{R.}~\bibnamefont{Winkler}},\ }%
  \bibfield{journal}{%
  \Doi{10.1103/PhysRevB.70.195322}{\bibinfo {journal} {Phys. Rev. B}}\ }%
  \textbf{\bibinfo {volume} {70}},\ \bibinfo {pages} {195322} (\bibinfo {year}
  {2004})%
  \bibAnnoteFile{NoStop}{Kainz2004}%
\bibitem{Hirsch1999}%
  \BibitemOpen
  \bibfield{author}{%
  \bibinfo {author} {\bibfnamefont{J.~E.}\ \bibnamefont{Hirsch}},\ }%
  \bibfield{journal}{%
  \Doi{10.1103/PhysRevLett.83.1834}{\bibinfo {journal} {Phys. Rev. Lett.}}\ }%
  \textbf{\bibinfo {volume} {83}},\ \bibinfo {pages} {1834} (\bibinfo {year}
  {1999})%
  \bibAnnoteFile{NoStop}{Hirsch1999}%
\bibitem{Sinova2004}%
  \BibitemOpen
  \bibfield{author}{%
  \bibinfo {author} {\bibfnamefont{J.}~\bibnamefont{Sinova}}, \bibinfo {author}
  {\bibfnamefont{D.}~\bibnamefont{Culcer}}, \bibinfo {author}
  {\bibfnamefont{Q.}~\bibnamefont{Niu}}, \bibinfo {author}
  {\bibfnamefont{N.~A.}\ \bibnamefont{Sinitsyn}}, \bibinfo {author}
  {\bibfnamefont{T.}~\bibnamefont{Jungwirth}},\ and\ \bibinfo {author}
  {\bibfnamefont{A.~H.}\ \bibnamefont{MacDonald}},\ }%
  \bibfield{journal}{%
  \Doi{10.1103/PhysRevLett.92.126603}{\bibinfo {journal} {Phys. Rev. Lett.}}\
  }%
  \textbf{\bibinfo {volume} {92}},\ \bibinfo {pages} {126603} (\bibinfo {year}
  {2004})%
  \bibAnnoteFile{NoStop}{Sinova2004}%
\bibitem{Kato2004}%
  \BibitemOpen
  \bibfield{author}{%
  \bibinfo {author} {\bibfnamefont{Y.~K.}\ \bibnamefont{Kato}}, \bibinfo
  {author} {\bibfnamefont{R.~C.}\ \bibnamefont{Myers}}, \bibinfo {author}
  {\bibfnamefont{A.~C.}\ \bibnamefont{Gossard}},\ and\ \bibinfo {author}
  {\bibfnamefont{D.}~\bibnamefont{Awschalom}},\ }%
  \bibfield{journal}{%
  \bibinfo {journal} {Science}\ }%
  \textbf{\bibinfo {volume} {1910}},\ \bibinfo {pages} {306} (\bibinfo {year}
  {2004})%
  \bibAnnoteFile{NoStop}{Kato2004}%
\bibitem{Koenig2007}%
  \BibitemOpen
  \bibfield{author}{%
  \bibinfo {author} {\bibfnamefont{M.}~\bibnamefont{Koenig}}, \bibinfo {author}
  {\bibfnamefont{S.}~\bibnamefont{Wiedmann}}, \bibinfo {author}
  {\bibfnamefont{C.}~\bibnamefont{Bruene}}, \bibinfo {author}
  {\bibfnamefont{A.}~\bibnamefont{Roth}}, \bibinfo {author}
  {\bibfnamefont{H.}~\bibnamefont{Buhmann}}, \bibinfo {author}
  {\bibfnamefont{L.~W.}\ \bibnamefont{Molenkamp}}, \bibinfo {author}
  {\bibfnamefont{X.-L.}\ \bibnamefont{Qi}},\ and\ \bibinfo {author}
  {\bibfnamefont{Z.}~\bibnamefont{S.-C.}},\ }%
  \bibfield{journal}{%
  \bibinfo {journal} {Science}\ }%
  \textbf{\bibinfo {volume} {318}},\ \bibinfo {pages} {766} (\bibinfo {year}
  {2007})%
  \bibAnnoteFile{NoStop}{Koenig2007}%
\bibitem{Bernevig2006}%
  \BibitemOpen
  \bibfield{author}{%
  \bibinfo {author} {\bibfnamefont{B.~A.}\ \bibnamefont{Bernevig}}, \bibinfo
  {author} {\bibfnamefont{J.}~\bibnamefont{Orenstein}},\ and\ \bibinfo {author}
  {\bibfnamefont{S.-C.}\ \bibnamefont{Zhang}},\ }%
  \bibfield{journal}{%
  \Doi{10.1103/PhysRevLett.97.236601}{\bibinfo {journal} {Phys. Rev. Lett.}}\
  }%
  \textbf{\bibinfo {volume} {97}},\ \bibinfo {pages} {236601} (\bibinfo {year}
  {2006})%
  \bibAnnoteFile{NoStop}{Bernevig2006}%
\bibitem{Koralek2009}%
  \BibitemOpen
  \bibfield{author}{%
  \bibinfo {author} {\bibfnamefont{J.~D.}\ \bibnamefont{Koralek}}, \bibinfo
  {author} {\bibfnamefont{C.~P.}\ \bibnamefont{Weber}}, \bibinfo {author}
  {\bibfnamefont{J.}~\bibnamefont{Orenstein}}, \bibinfo {author}
  {\bibfnamefont{B.~A.}\ \bibnamefont{Bernevig}}, \bibinfo {author}
  {\bibfnamefont{S.-C.}\ \bibnamefont{Zhang}}, \bibinfo {author}
  {\bibfnamefont{S.}~\bibnamefont{Mack}},\ and\ \bibinfo {author}
  {\bibfnamefont{D.~D.}\ \bibnamefont{Awschalom}},\ }%
  \bibfield{journal}{%
  \Doi{10.1038/nature07871}{\bibinfo {journal} {Nature}}\ }%
  \textbf{\bibinfo {volume} {458}},\ \bibinfo {pages} {610} (\bibinfo {year}
  {2009})%
  \bibAnnoteFile{NoStop}{Koralek2009}%
\bibitem{Walser2012}%
  \BibitemOpen
  \bibfield{author}{%
  \bibinfo {author} {\bibfnamefont{M.~P.}\ \bibnamefont{Walser}}, \bibinfo
  {author} {\bibfnamefont{C.}~\bibnamefont{Reichl}}, \bibinfo {author}
  {\bibfnamefont{W.}~\bibnamefont{Wegscheider}},\ and\ \bibinfo {author}
  {\bibfnamefont{G.}~\bibnamefont{Salis}},\ }%
  \bibfield{journal}{%
  \Doi{10.1038/nphys2383}{\bibinfo {journal} {Nature Phys.}}\ }%
  \textbf{\bibinfo {volume} {8}},\ \bibinfo {pages} {757} (\bibinfo {year}
  {2012})%
  \bibAnnoteFile{NoStop}{Walser2012}%
\bibitem{Mourik2012}%
  \BibitemOpen
  \bibfield{author}{%
  \bibinfo {author} {\bibfnamefont{V.}~\bibnamefont{Mourik}}, \bibinfo {author}
  {\bibfnamefont{K.}~\bibnamefont{Zuo}}, \bibinfo {author}
  {\bibfnamefont{S.~M.}\ \bibnamefont{Frolov}}, \bibinfo {author}
  {\bibfnamefont{S.~R.}\ \bibnamefont{Plissard}}, \bibinfo {author}
  {\bibfnamefont{E.~P. A.~M.}\ \bibnamefont{Bakkers}},\ and\ \bibinfo {author}
  {\bibfnamefont{L.~P.}\ \bibnamefont{Kouwenhoven}},\ }%
  \bibfield{journal}{%
  \Doi{10.1126/science.1222360}{\bibinfo {journal} {Science}}\ }%
  \textbf{\bibinfo {volume} {336}},\ \bibinfo {pages} {1003} (\bibinfo {year}
  {2012})%
  \bibAnnoteFile{NoStop}{Mourik2012}%
\bibitem{Debray2009}%
  \BibitemOpen
  \bibfield{author}{%
  \bibinfo {author} {\bibfnamefont{P.}~\bibnamefont{Debray}}, \bibinfo {author}
  {\bibfnamefont{S.~M.~S.}\ \bibnamefont{Rahman}}, \bibinfo {author}
  {\bibfnamefont{J.}~\bibnamefont{Wan}}, \bibinfo {author}
  {\bibfnamefont{R.~S.}\ \bibnamefont{Newrock}}, \bibinfo {author}
  {\bibfnamefont{M.}~\bibnamefont{Cahay}}, \bibinfo {author}
  {\bibfnamefont{A.~T.}\ \bibnamefont{Ngo}}, \bibinfo {author}
  {\bibfnamefont{S.~E.}\ \bibnamefont{Ulloa}}, \bibinfo {author}
  {\bibfnamefont{S.~T.}\ \bibnamefont{Herbert}}, \bibinfo {author}
  {\bibfnamefont{M.}~\bibnamefont{Muhammad}},\ and\ \bibinfo {author}
  {\bibfnamefont{J.}~\bibnamefont{M.}},\ }%
  \bibfield{journal}{%
  \Doi{10.1038/nnano.2009.240}{\bibinfo {journal} {Nature Nanotech.}}\ }%
  \textbf{\bibinfo {volume} {4}},\ \bibinfo {pages} {759} (\bibinfo {year}
  {2009})%
  \bibAnnoteFile{NoStop}{Debray2009}%
\bibitem{Datta1990}%
  \BibitemOpen
  \bibfield{author}{%
  \bibinfo {author} {\bibfnamefont{S.}~\bibnamefont{Datta}}\ and\ \bibinfo
  {author} {\bibfnamefont{B.}~\bibnamefont{Das}},\ }%
  \bibfield{journal}{%
  \Doi{http://dx.doi.org/10.1063/1.102730}{\bibinfo {journal} {Appl. Phys.
  Lett.}}\ }%
  \textbf{\bibinfo {volume} {56}},\ \bibinfo {pages} {665} (\bibinfo {year}
  {1990})%
  \bibAnnoteFile{NoStop}{Datta1990}%
\bibitem{Schliemann2003}%
  \BibitemOpen
  \bibfield{author}{%
  \bibinfo {author} {\bibfnamefont{J.}~\bibnamefont{Schliemann}}, \bibinfo
  {author} {\bibfnamefont{J.~C.}\ \bibnamefont{Egues}},\ and\ \bibinfo {author}
  {\bibfnamefont{D.}~\bibnamefont{Loss}},\ }%
  \bibfield{journal}{%
  \Doi{10.1103/PhysRevLett.90.146801}{\bibinfo {journal} {Phys. Rev. Lett.}}\
  }%
  \textbf{\bibinfo {volume} {90}},\ \bibinfo {pages} {146801} (\bibinfo {year}
  {2003})%
  \bibAnnoteFile{NoStop}{Schliemann2003}%
\bibitem{Zutic2004}%
  \BibitemOpen
  \bibfield{author}{%
  \bibinfo {author} {\bibfnamefont{I.}~\bibnamefont{\ifmmode \check{Z}\else
  \v{Z}\fi{}uti\ifmmode~\acute{c}\else \'{c}\fi{}}}, \bibinfo {author}
  {\bibfnamefont{J.}~\bibnamefont{Fabian}},\ and\ \bibinfo {author}
  {\bibfnamefont{S.}~\bibnamefont{Das~Sarma}},\ }%
  \bibfield{journal}{%
  \Doi{10.1103/RevModPhys.76.323}{\bibinfo {journal} {Rev. Mod. Phys.}}\ }%
  \textbf{\bibinfo {volume} {76}},\ \bibinfo {pages} {323} (\bibinfo {year}
  {2004})%
  \bibAnnoteFile{NoStop}{Zutic2004}%
\bibitem{Chuang2015}%
  \BibitemOpen
  \bibfield{author}{%
  \bibinfo {author} {\bibfnamefont{P.}~\bibnamefont{Chuang}}, \bibinfo {author}
  {\bibfnamefont{S.-C.}\ \bibnamefont{Ho}}, \bibinfo {author}
  {\bibfnamefont{L.~W.}\ \bibnamefont{Smith}}, \bibinfo {author}
  {\bibfnamefont{F.}~\bibnamefont{Sfigakis}}, \bibinfo {author}
  {\bibfnamefont{M.}~\bibnamefont{Pepper}}, \bibinfo {author}
  {\bibfnamefont{C.-H.}\ \bibnamefont{Chen}}, \bibinfo {author}
  {\bibfnamefont{J.-C.}\ \bibnamefont{Fan}}, \bibinfo {author}
  {\bibfnamefont{J.~P.}\ \bibnamefont{Griffiths}}, \bibinfo {author}
  {\bibfnamefont{I.}~\bibnamefont{Farrer}}, \bibinfo {author}
  {\bibfnamefont{H.~E.}\ \bibnamefont{Beere}}, \bibinfo {author}
  {\bibfnamefont{G.~A.~C.}\ \bibnamefont{Jones}}, \bibinfo {author}
  {\bibfnamefont{D.~A.}\ \bibnamefont{Ritchie}},\ and\ \bibinfo {author}
  {\bibfnamefont{T.-M.}\ \bibnamefont{Chen}},\ }%
  \bibfield{journal}{%
  \bibinfo {journal} {Nat Nano}\ }%
  \textbf{\bibinfo {volume} {10}},\ \bibinfo {pages} {35} (\bibinfo {year}
  {2015})%
  \bibAnnoteFile{NoStop}{Chuang2015}%
\bibitem{Bednarek2008}%
  \BibitemOpen
  \bibfield{author}{%
  \bibinfo {author} {\bibfnamefont{S.}~\bibnamefont{Bednarek}}\ and\ \bibinfo
  {author} {\bibfnamefont{B.}~\bibnamefont{Szafran}},\ }%
  \bibfield{journal}{%
  \Doi{10.1103/PhysRevLett.101.216805}{\bibinfo {journal} {Phys. Rev. Lett.}}\
  }%
  \textbf{\bibinfo {volume} {101}},\ \bibinfo {pages} {216805} (\bibinfo {year}
  {2008})%
  \bibAnnoteFile{NoStop}{Bednarek2008}%
\bibitem{Meier}%
  \BibitemOpen
  \bibfield{author}{%
  \bibinfo {author} {\bibfnamefont{L.}~\bibnamefont{Meier}}, \bibinfo {author}
  {\bibfnamefont{G.}~\bibnamefont{Salis}}, \bibinfo {author}
  {\bibfnamefont{I.}~\bibnamefont{Shorubalko}}, \bibinfo {author}
  {\bibfnamefont{E.}~\bibnamefont{Gini}}, \bibinfo {author}
  {\bibfnamefont{S.}~\bibnamefont{Schoen}},\ and\ \bibinfo {author}
  {\bibfnamefont{K.}~\bibnamefont{Ensslin}},\ }%
  \bibfield{journal}{%
  \bibinfo {journal} {Nature Phys.}\ }%
  \textbf{\bibinfo {volume} {3}} (\bibinfo {year} {2007})%
  \bibAnnoteFile{NoStop}{Meier}%
\bibitem{Bychkov1984}%
  \BibitemOpen
  \bibfield{author}{%
  \bibinfo {author} {\bibfnamefont{Y.}~\bibnamefont{Bychkov}}\ and\ \bibinfo
  {author} {\bibfnamefont{E.}~\bibnamefont{Rashba}},\ }%
  \bibfield{journal}{%
  \bibinfo {journal} {J. Phys. C}\ }%
  \textbf{\bibinfo {volume} {17}},\ \bibinfo {pages} {6039} (\bibinfo {year}
  {1984})%
  \bibAnnoteFile{NoStop}{Bychkov1984}%
\bibitem{Nitta1997}%
  \BibitemOpen
  \bibfield{author}{%
  \bibinfo {author} {\bibfnamefont{J.}~\bibnamefont{Nitta}}, \bibinfo {author}
  {\bibfnamefont{T.}~\bibnamefont{Akazaki}}, \bibinfo {author}
  {\bibfnamefont{H.}~\bibnamefont{Takayanagi}},\ and\ \bibinfo {author}
  {\bibfnamefont{T.}~\bibnamefont{Enoki}},\ }%
  \bibfield{journal}{%
  \Doi{10.1103/PhysRevLett.78.1335}{\bibinfo {journal} {Phys. Rev. Lett.}}\ }%
  \textbf{\bibinfo {volume} {78}},\ \bibinfo {pages} {1335} (\bibinfo {year}
  {1997})%
  \bibAnnoteFile{NoStop}{Nitta1997}%
\bibitem{Engels1997}%
  \BibitemOpen
  \bibfield{author}{%
  \bibinfo {author} {\bibfnamefont{G.}~\bibnamefont{Engels}}, \bibinfo {author}
  {\bibfnamefont{J.}~\bibnamefont{Lange}}, \bibinfo {author}
  {\bibfnamefont{T.}~\bibnamefont{Sch\"apers}},\ and\ \bibinfo {author}
  {\bibfnamefont{H.}~\bibnamefont{L\"uth}},\ }%
  \bibfield{journal}{%
  \Doi{10.1103/PhysRevB.55.R1958}{\bibinfo {journal} {Phys. Rev. B}}\ }%
  \textbf{\bibinfo {volume} {55}},\ \bibinfo {pages} {R1958} (\bibinfo {year}
  {1997})%
  \bibAnnoteFile{NoStop}{Engels1997}%
\bibitem{Lo2002}%
  \BibitemOpen
  \bibfield{author}{%
  \bibinfo {author} {\bibfnamefont{I.}~\bibnamefont{Lo}}, \bibinfo {author}
  {\bibfnamefont{J.~K.}\ \bibnamefont{Tsai}}, \bibinfo {author}
  {\bibfnamefont{W.~J.}\ \bibnamefont{Yao}}, \bibinfo {author}
  {\bibfnamefont{P.~C.}\ \bibnamefont{Ho}}, \bibinfo {author}
  {\bibfnamefont{L.~W.}\ \bibnamefont{Tu}}, \bibinfo {author}
  {\bibfnamefont{T.~C.}\ \bibnamefont{Chang}}, \bibinfo {author}
  {\bibfnamefont{S.}~\bibnamefont{Elhamri}}, \bibinfo {author}
  {\bibfnamefont{W.~C.}\ \bibnamefont{Mitchel}}, \bibinfo {author}
  {\bibfnamefont{K.~Y.}\ \bibnamefont{Hsieh}}, \bibinfo {author}
  {\bibfnamefont{J.~H.}\ \bibnamefont{Huang}}, \bibinfo {author}
  {\bibfnamefont{H.~L.}\ \bibnamefont{Huang}}, ,\ and\ \bibinfo {author}
  {\bibfnamefont{W.-C.}\ \bibnamefont{Tsai}},\ }%
  \bibfield{journal}{%
  \bibinfo {journal} {Phys. Rev. B}\ }%
  \textbf{\bibinfo {volume} {65}},\ \bibinfo {pages} {R161306} (\bibinfo {year}
  {2002})%
  \bibAnnoteFile{NoStop}{Lo2002}%
\bibitem{Kwon2007}%
  \BibitemOpen
  \bibfield{author}{%
  \bibinfo {author} {\bibfnamefont{J.~H.}\ \bibnamefont{Kwon}}, \bibinfo
  {author} {\bibfnamefont{H.~C.}\ \bibnamefont{Koo}}, \bibinfo {author}
  {\bibfnamefont{J.}~\bibnamefont{Chang}}, \bibinfo {author}
  {\bibfnamefont{S.-H.}\ \bibnamefont{Han}},\ and\ \bibinfo {author}
  {\bibfnamefont{J.}~\bibnamefont{Eom}},\ }%
  \bibfield{journal}{%
  \Doi{http://dx.doi.org/10.1063/1.2714993}{\bibinfo {journal} {Appl. Phys.
  Lett.}}\ }%
  \textbf{\bibinfo {volume} {90}},\ \bibinfo {eid} {112505} (\bibinfo {year}
  {2007})%
  \bibAnnoteFile{NoStop}{Kwon2007}%
\bibitem{Grundler2000}%
  \BibitemOpen
  \bibfield{author}{%
  \bibinfo {author} {\bibfnamefont{D.}~\bibnamefont{Grundler}},\ }%
  \bibfield{journal}{%
  \Doi{10.1103/PhysRevLett.84.6074}{\bibinfo {journal} {Phys. Rev. Lett.}}\ }%
  \textbf{\bibinfo {volume} {84}},\ \bibinfo {pages} {6074} (\bibinfo {year}
  {2000})%
  \bibAnnoteFile{NoStop}{Grundler2000}%
\bibitem{Kim2010}%
  \BibitemOpen
  \bibfield{author}{%
  \bibinfo {author} {\bibfnamefont{K.-H.}\ \bibnamefont{Kim}}, \bibinfo
  {author} {\bibfnamefont{H.-j.}\ \bibnamefont{Kim}}, \bibinfo {author}
  {\bibfnamefont{H.~C.}\ \bibnamefont{Koo}}, \bibinfo {author}
  {\bibfnamefont{J.}~\bibnamefont{Chang}},\ and\ \bibinfo {author}
  {\bibfnamefont{S.-H.}\ \bibnamefont{Han}},\ }%
  \bibfield{journal}{%
  \Doi{http://dx.doi.org/10.1063/1.3462325}{\bibinfo {journal} {Appl. Phys.
  Lett.}}\ }%
  \textbf{\bibinfo {volume} {97}},\ \bibinfo {eid} {012504} (\bibinfo {year}
  {2010})%
  \bibAnnoteFile{NoStop}{Kim2010}%
\bibitem{Das1989}%
  \BibitemOpen
  \bibfield{author}{%
  \bibinfo {author} {\bibfnamefont{B.}~\bibnamefont{Das}}, \bibinfo {author}
  {\bibfnamefont{D.~C.}\ \bibnamefont{Miller}}, \bibinfo {author}
  {\bibfnamefont{S.}~\bibnamefont{Datta}}, \bibinfo {author}
  {\bibfnamefont{R.}~\bibnamefont{Reifenberger}}, \bibinfo {author}
  {\bibfnamefont{W.~P.}\ \bibnamefont{Hong}}, \bibinfo {author}
  {\bibfnamefont{P.~K.}\ \bibnamefont{Bhattacharya}}, \bibinfo {author}
  {\bibfnamefont{J.}~\bibnamefont{Singh}},\ and\ \bibinfo {author}
  {\bibfnamefont{M.}~\bibnamefont{Jaffe}},\ }%
  \bibfield{journal}{%
  \Doi{10.1103/PhysRevB.39.1411}{\bibinfo {journal} {Phys. Rev. B}}\ }%
  \textbf{\bibinfo {volume} {39}},\ \bibinfo {pages} {1411} (\bibinfo {year}
  {1989})%
  \bibAnnoteFile{NoStop}{Das1989}%
\bibitem{Park2013}%
  \BibitemOpen
  \bibfield{author}{%
  \bibinfo {author} {\bibfnamefont{Y.}~\bibnamefont{Ho~Park}}, \bibinfo
  {author} {\bibfnamefont{H.-j.}\ \bibnamefont{Kim}}, \bibinfo {author}
  {\bibfnamefont{J.}~\bibnamefont{Chang}}, \bibinfo {author}
  {\bibfnamefont{S.}~\bibnamefont{Hee~Han}}, \bibinfo {author}
  {\bibfnamefont{J.}~\bibnamefont{Eom}}, \bibinfo {author}
  {\bibfnamefont{H.-J.}\ \bibnamefont{Choi}},\ and\ \bibinfo {author}
  {\bibfnamefont{H.}~\bibnamefont{Cheol~Koo}},\ }%
  \bibfield{journal}{%
  \Doi{http://dx.doi.org/10.1063/1.4855495}{\bibinfo {journal} {Appl. Phys.
  Lett.}}\ }%
  \textbf{\bibinfo {volume} {103}},\ \bibinfo {eid} {252407} (\bibinfo {year}
  {2013})%
  \bibAnnoteFile{NoStop}{Park2013}%
\bibitem{Koga2002}%
  \BibitemOpen
  \bibfield{author}{%
  \bibinfo {author} {\bibfnamefont{T.}~\bibnamefont{Koga}}, \bibinfo {author}
  {\bibfnamefont{J.}~\bibnamefont{Nitta}}, \bibinfo {author}
  {\bibfnamefont{T.}~\bibnamefont{Akazaki}},\ and\ \bibinfo {author}
  {\bibfnamefont{H.}~\bibnamefont{Takayanagi}},\ }%
  \bibfield{journal}{%
  \Doi{10.1103/PhysRevLett.89.046801}{\bibinfo {journal} {Phys. Rev. Lett.}}\
  }%
  \textbf{\bibinfo {volume} {89}},\ \bibinfo {pages} {046801} (\bibinfo {year}
  {2002})%
  \bibAnnoteFile{NoStop}{Koga2002}%
\bibitem{Ganichev2004}%
  \BibitemOpen
  \bibfield{author}{%
  \bibinfo {author} {\bibfnamefont{S.~D.}\ \bibnamefont{Ganichev}}, \bibinfo
  {author} {\bibfnamefont{V.~V.}\ \bibnamefont{Bel'kov}}, \bibinfo {author}
  {\bibfnamefont{L.~E.}\ \bibnamefont{Golub}}, \bibinfo {author}
  {\bibfnamefont{E.~L.}\ \bibnamefont{Ivchenko}}, \bibinfo {author}
  {\bibfnamefont{P.}~\bibnamefont{Schneider}}, \bibinfo {author}
  {\bibfnamefont{S.}~\bibnamefont{Giglberger}}, \bibinfo {author}
  {\bibfnamefont{J.}~\bibnamefont{Eroms}}, \bibinfo {author}
  {\bibfnamefont{J.}~\bibnamefont{De~Boeck}}, \bibinfo {author}
  {\bibfnamefont{G.}~\bibnamefont{Borghs}}, \bibinfo {author}
  {\bibfnamefont{W.}~\bibnamefont{Wegscheider}}, \bibinfo {author}
  {\bibfnamefont{D.}~\bibnamefont{Weiss}},\ and\ \bibinfo {author}
  {\bibfnamefont{W.}~\bibnamefont{Prettl}},\ }%
  \bibfield{journal}{%
  \Doi{10.1103/PhysRevLett.92.256601}{\bibinfo {journal} {Phys. Rev. Lett.}}\
  }%
  \textbf{\bibinfo {volume} {92}},\ \bibinfo {pages} {256601} (\bibinfo {year}
  {2004})%
  \bibAnnoteFile{NoStop}{Ganichev2004}%
\bibitem{Meier2007}%
  \BibitemOpen
  \bibfield{author}{%
  \bibinfo {author} {\bibfnamefont{L.}~\bibnamefont{Meier}}, \bibinfo {author}
  {\bibfnamefont{G.}~\bibnamefont{Salis}}, \bibinfo {author}
  {\bibfnamefont{I.}~\bibnamefont{Shorubalko}}, \bibinfo {author}
  {\bibfnamefont{E.}~\bibnamefont{Gini}}, \bibinfo {author}
  {\bibfnamefont{S.}~\bibnamefont{Schon}},\ and\ \bibinfo {author}
  {\bibfnamefont{K.}~\bibnamefont{Ensslin}},\ }%
  \bibfield{journal}{%
  \Doi{10.1038/nphys675}{\bibinfo {journal} {Nature Phys.}}\ }%
  \textbf{\bibinfo {volume} {3}},\ \bibinfo {pages} {650} (\bibinfo {year}
  {2007})%
  \bibAnnoteFile{NoStop}{Meier2007}%
\bibitem{Sellier2011}%
  \BibitemOpen
  \bibfield{author}{%
  \bibinfo {author} {\bibfnamefont{H.}~\bibnamefont{Sellier}}, \bibinfo
  {author} {\bibfnamefont{B.}~\bibnamefont{Hackens}}, \bibinfo {author}
  {\bibfnamefont{M.~G.}\ \bibnamefont{Pala}}, \bibinfo {author}
  {\bibfnamefont{F.}~\bibnamefont{Martins}}, \bibinfo {author}
  {\bibfnamefont{S.}~\bibnamefont{Baltazar}}, \bibinfo {author}
  {\bibfnamefont{X.}~\bibnamefont{Wallart}}, \bibinfo {author}
  {\bibfnamefont{L.}~\bibnamefont{Desplanque}}, \bibinfo {author}
  {\bibfnamefont{V.}~\bibnamefont{Bayot}},\ and\ \bibinfo {author}
  {\bibfnamefont{S.}~\bibnamefont{Huant}},\ }%
  \bibfield{journal}{%
  \bibinfo {journal} {Semicond. Sci. Technol.}\ }%
  \textbf{\bibinfo {volume} {26}},\ \bibinfo {pages} {064008} (\bibinfo {year}
  {2011})%
  \bibAnnoteFile{NoStop}{Sellier2011}%
\bibitem{Ferry2011}%
  \BibitemOpen
  \bibfield{author}{%
  \bibinfo {author} {\bibfnamefont{D.~K.}\ \bibnamefont{Ferry}}, \bibinfo
  {author} {\bibfnamefont{A.~M.}\ \bibnamefont{Burke}}, \bibinfo {author}
  {\bibfnamefont{R.}~\bibnamefont{Akis}}, \bibinfo {author}
  {\bibfnamefont{R.}~\bibnamefont{Brunner}}, \bibinfo {author}
  {\bibfnamefont{T.~E.}\ \bibnamefont{Day}}, \bibinfo {author}
  {\bibfnamefont{R.}~\bibnamefont{Meisels}}, \bibinfo {author}
  {\bibfnamefont{F.}~\bibnamefont{Kuchar}}, \bibinfo {author}
  {\bibfnamefont{J.~P.}\ \bibnamefont{Bird}},\ and\ \bibinfo {author}
  {\bibfnamefont{B.~R.}\ \bibnamefont{Bennett}},\ }%
  \bibfield{journal}{%
  \bibinfo {journal} {Sem. Sci. Tech.}\ }%
  \textbf{\bibinfo {volume} {26}},\ \bibinfo {pages} {043001} (\bibinfo {year}
  {2011})%
  \bibAnnoteFile{NoStop}{Ferry2011}%
\bibitem{Topinka2001}%
  \BibitemOpen
  \bibfield{author}{%
  \bibinfo {author} {\bibfnamefont{M.~A.}\ \bibnamefont{Topinka}}, \bibinfo
  {author} {\bibfnamefont{B.~J.}\ \bibnamefont{LeRoy}}, \bibinfo {author}
  {\bibfnamefont{R.~M.}\ \bibnamefont{Westervelt}}, \bibinfo {author}
  {\bibfnamefont{S.~E.~J.}\ \bibnamefont{Shaw}}, \bibinfo {author}
  {\bibfnamefont{R.}~\bibnamefont{Fleischmann}}, \bibinfo {author}
  {\bibfnamefont{E.~J.}\ \bibnamefont{Heller}}, \bibinfo {author}
  {\bibfnamefont{K.~D.}\ \bibnamefont{Maranowski}},\ and\ \bibinfo {author}
  {\bibfnamefont{A.~C.}\ \bibnamefont{Gossard}},\ }%
  \bibfield{journal}{%
  \Doi{10.1038/35065553}{\bibinfo {journal} {Nature}}\ }%
  \textbf{\bibinfo {volume} {410}},\ \bibinfo {pages} {183} (\bibinfo {year}
  {2001})%
  \bibAnnoteFile{NoStop}{Topinka2001}%
\bibitem{Schnez2011}%
  \BibitemOpen
  \bibfield{author}{%
  \bibinfo {author} {\bibfnamefont{S.}~\bibnamefont{Schnez}}, \bibinfo {author}
  {\bibfnamefont{C.}~\bibnamefont{R\"ossler}}, \bibinfo {author}
  {\bibfnamefont{T.}~\bibnamefont{Ihn}}, \bibinfo {author}
  {\bibfnamefont{K.}~\bibnamefont{Ensslin}}, \bibinfo {author}
  {\bibfnamefont{C.}~\bibnamefont{Reichl}},\ and\ \bibinfo {author}
  {\bibfnamefont{W.}~\bibnamefont{Wegscheider}},\ }%
  \bibfield{journal}{%
  \Doi{10.1103/PhysRevB.84.195322}{\bibinfo {journal} {Phys. Rev. B}}\ }%
  \textbf{\bibinfo {volume} {84}},\ \bibinfo {pages} {195322} (\bibinfo {year}
  {2011})%
  \bibAnnoteFile{NoStop}{Schnez2011}%
\bibitem{Jura2007}%
  \BibitemOpen
  \bibfield{author}{%
  \bibinfo {author} {\bibfnamefont{M.~P.}\ \bibnamefont{Jura}}, \bibinfo
  {author} {\bibfnamefont{M.~A.}\ \bibnamefont{Topinka}}, \bibinfo {author}
  {\bibfnamefont{L.}~\bibnamefont{Urban}}, \bibinfo {author}
  {\bibfnamefont{A.}~\bibnamefont{Yazdani}}, \bibinfo {author}
  {\bibfnamefont{H.}~\bibnamefont{Shtrikman}}, \bibinfo {author}
  {\bibfnamefont{L.~N.}\ \bibnamefont{Pfeiffer}}, \bibinfo {author}
  {\bibfnamefont{K.~W.}\ \bibnamefont{West}},\ and\ \bibinfo {author}
  {\bibfnamefont{D.}~\bibnamefont{Goldhaber-Gordon}},\ }%
  \bibfield{journal}{%
  \Doi{10.1038/nphys756}{\bibinfo {journal} {Nature Phys.}}\ }%
  \textbf{\bibinfo {volume} {3}},\ \bibinfo {pages} {841} (\bibinfo {year}
  {2007})%
  \bibAnnoteFile{NoStop}{Jura2007}%
\bibitem{Jura2009}%
  \BibitemOpen
  \bibfield{author}{%
  \bibinfo {author} {\bibfnamefont{M.~P.}\ \bibnamefont{Jura}}, \bibinfo
  {author} {\bibfnamefont{M.~A.}\ \bibnamefont{Topinka}}, \bibinfo {author}
  {\bibfnamefont{M.}~\bibnamefont{Grobis}}, \bibinfo {author}
  {\bibfnamefont{L.~N.}\ \bibnamefont{Pfeiffer}}, \bibinfo {author}
  {\bibfnamefont{K.~W.}\ \bibnamefont{West}},\ and\ \bibinfo {author}
  {\bibfnamefont{D.}~\bibnamefont{Goldhaber-Gordon}},\ }%
  \bibfield{journal}{%
  \Doi{10.1103/PhysRevB.80.041303}{\bibinfo {journal} {Phys. Rev. B}}\ }%
  \textbf{\bibinfo {volume} {80}},\ \bibinfo {pages} {041303} (\bibinfo {year}
  {2009})%
  \bibAnnoteFile{NoStop}{Jura2009}%
\bibitem{Kleshchonok2015}%
  \BibitemOpen
  \bibfield{author}{%
  \bibinfo {author} {\bibfnamefont{A.}~\bibnamefont{Kleshchonok}}, \bibinfo
  {author} {\bibfnamefont{G.}~\bibnamefont{Fleury}}, \bibinfo {author}
  {\bibfnamefont{J.-L.}\ \bibnamefont{Pichard}},\ and\ \bibinfo {author}
  {\bibfnamefont{G.}~\bibnamefont{Lemari\'e}},\ }%
  \bibfield{journal}{%
  \Doi{10.1103/PhysRevB.91.125416}{\bibinfo {journal} {Phys. Rev. B}}\ }%
  \textbf{\bibinfo {volume} {91}},\ \bibinfo {pages} {125416} (\bibinfo {year}
  {2015})%
  \bibAnnoteFile{NoStop}{Kleshchonok2015}%
\bibitem{Topinka2000}%
  \BibitemOpen
  \bibfield{author}{%
  \bibinfo {author} {\bibfnamefont{M.~A.}\ \bibnamefont{Topinka}}, \bibinfo
  {author} {\bibfnamefont{B.~J.}\ \bibnamefont{LeRoy}}, \bibinfo {author}
  {\bibfnamefont{S.~E.~J.}\ \bibnamefont{Shaw}}, \bibinfo {author}
  {\bibfnamefont{E.~J.}\ \bibnamefont{Heller}}, \bibinfo {author}
  {\bibfnamefont{R.~M.}\ \bibnamefont{Westervelt}}, \bibinfo {author}
  {\bibfnamefont{K.~D.}\ \bibnamefont{Maranowski}},\ and\ \bibinfo {author}
  {\bibfnamefont{A.~C.}\ \bibnamefont{Gossard}},\ }%
  \bibfield{journal}{%
  \Doi{10.1126/science.289.5488.2323}{\bibinfo {journal} {Science}}\ }%
  \textbf{\bibinfo {volume} {289}},\ \bibinfo {pages} {2323} (\bibinfo {year}
  {2000})%
  \bibAnnoteFile{NoStop}{Topinka2000}%
\bibitem{Kozikov2015}%
  \BibitemOpen
  \bibfield{author}{%
  \bibinfo {author} {\bibfnamefont{A.~A.}\ \bibnamefont{Kozikov}}, \bibinfo
  {author} {\bibfnamefont{R.}~\bibnamefont{Steinacher}}, \bibinfo {author}
  {\bibfnamefont{C.}~\bibnamefont{R\"ossler}}, \bibinfo {author}
  {\bibfnamefont{T.}~\bibnamefont{Ihn}}, \bibinfo {author}
  {\bibfnamefont{K.}~\bibnamefont{Ensslin}}, \bibinfo {author}
  {\bibfnamefont{C.}~\bibnamefont{Reichl}},\ and\ \bibinfo {author}
  {\bibfnamefont{W.}~\bibnamefont{Wegscheider}},\ }%
  \bibfield{journal}{%
  \Doi{10.1021/acs.nanolett.5b03170}{\bibinfo {journal} {Nano Lett.}}\ }%
  \textbf{\bibinfo {volume} {15}},\ \bibinfo {pages} {7994} (\bibinfo {year}
  {2015})%
  \bibAnnoteFile{NoStop}{Kozikov2015}%
\bibitem{Nowak2014}%
  \BibitemOpen
  \bibfield{author}{%
  \bibinfo {author} {\bibfnamefont{M.~P.}\ \bibnamefont{Nowak}}, \bibinfo
  {author} {\bibfnamefont{K.}~\bibnamefont{Kolasi\ifmmode~\acute{n}\else
  \'{n}\fi{}ski}},\ and\ \bibinfo {author}
  {\bibfnamefont{B.}~\bibnamefont{Szafran}},\ }%
  \bibfield{journal}{%
  \Doi{10.1103/PhysRevB.90.035301}{\bibinfo {journal} {Phys. Rev. B}}\ }%
  \textbf{\bibinfo {volume} {90}},\ \bibinfo {pages} {035301} (\bibinfo {year}
  {2014})%
  \bibAnnoteFile{NoStop}{Nowak2014}%
\bibitem{Davies1995}%
  \BibitemOpen
  \bibfield{author}{%
  \bibinfo {author} {\bibfnamefont{J.~H.}\ \bibnamefont{Davies}}, \bibinfo
  {author} {\bibfnamefont{I.~A.}\ \bibnamefont{Larkin}},\ and\ \bibinfo
  {author} {\bibfnamefont{E.~V.}\ \bibnamefont{Sukhorukov}},\ }%
  \bibfield{journal}{%
  \bibinfo {journal} {J. Appl. Phys}\ }%
  \textbf{\bibinfo {volume} {77}},\ \bibinfo {pages} {4504} (\bibinfo {year}
  {1995}),\ \bibinfo {note} {we apply the formula for the finite rectangle
  gate, given by equation $V_\mathrm{QPC}/V_\mathrm{g} = g(x-L,y-B) +
  g(x-L,T-y) + g(R-x,y-B) + g(R-x,T-y)$, where $g(u,v) =
  \frac{1}{2\pi}\arctan{\frac{uv}{dR}};~R=\sqrt{v^2+u^2+d^2}$, with $L=100$nm,
  $R=300$nm, $B=-100$nm, $T=1300$nm and $d=50$nm.}%
  \bibAnnoteFile{Stop}{Davies1995}%
\bibitem{kolasinskiDFT2013}%
  \BibitemOpen
  \bibfield{author}{%
  \bibinfo {author}
  {\bibfnamefont{K.}~\bibnamefont{Kolasi\ifmmode~\acute{n}\else
  \'{n}\fi{}ski}}\ and\ \bibinfo {author}
  {\bibfnamefont{B.}~\bibnamefont{Szafran}},\ }%
  \bibfield{journal}{%
  \Doi{10.1103/PhysRevB.88.165306}{\bibinfo {journal} {Phys. Rev. B}}\ }%
  \textbf{\bibinfo {volume} {88}},\ \bibinfo {pages} {165306} (\bibinfo {year}
  {2013})%
  \bibAnnoteFile{NoStop}{kolasinskiDFT2013}%
\bibitem{Steinacher2015}%
  \BibitemOpen
  \bibfield{author}{%
  \bibinfo {author} {\bibfnamefont{R.}~\bibnamefont{Steinacher}}, \bibinfo
  {author} {\bibfnamefont{A.~A.}\ \bibnamefont{Kozikov}}, \bibinfo {author}
  {\bibfnamefont{C.}~\bibnamefont{R\"ossler}}, \bibinfo {author}
  {\bibfnamefont{C.}~\bibnamefont{Reichl}}, \bibinfo {author}
  {\bibfnamefont{W.}~\bibnamefont{Wegscheider}}, \bibinfo {author}
  {\bibfnamefont{T.}~\bibnamefont{Ihn}},\ and\ \bibinfo {author}
  {\bibfnamefont{K.}~\bibnamefont{Ensslin}},\ }%
  \bibfield{journal}{%
  \bibinfo {journal} {New J. Phys.}\ }%
  \textbf{\bibinfo {volume} {17}},\ \bibinfo {pages} {043043} (\bibinfo {year}
  {2015})%
  \bibAnnoteFile{NoStop}{Steinacher2015}%
\bibitem{Bhandari2013}%
  \BibitemOpen
  \bibfield{author}{%
  \bibinfo {author} {\bibfnamefont{N.}~\bibnamefont{Bhandari}}, \bibinfo
  {author} {\bibfnamefont{M.}~\bibnamefont{Dutta}}, \bibinfo {author}
  {\bibfnamefont{J.}~\bibnamefont{Charles}}, \bibinfo {author}
  {\bibfnamefont{R.~S.}\ \bibnamefont{Newrock}}, \bibinfo {author}
  {\bibfnamefont{M.}~\bibnamefont{Cahay}},\ and\ \bibinfo {author}
  {\bibfnamefont{S.~T.}\ \bibnamefont{Herbert}},\ }%
  \bibfield{journal}{%
  \bibinfo {journal} {Adv. Nat. Sci: Nanosci. Nanotechnol.}\ }%
  \textbf{\bibinfo {volume} {4}},\ \bibinfo {pages} {013002} (\bibinfo {year}
  {2013})%
  \bibAnnoteFile{NoStop}{Bhandari2013}%
\bibitem{Kolasinski2016Lande}%
  \BibitemOpen
  \bibfield{author}{%
  \bibinfo {author}
  {\bibfnamefont{K.}~\bibnamefont{Kolasi\ifmmode~\acute{n}\else
  \'{n}\fi{}ski}}, \bibinfo {author}
  {\bibfnamefont{A.}~\bibnamefont{Mre\ifmmode \acute{n}\else
  \'{n}\fi{}ca-Kolasi\ifmmode~\acute{n}\else \'{n}\fi{}ska}},\ and\ \bibinfo
  {author} {\bibfnamefont{B.}~\bibnamefont{Szafran}},\ }%
  \bibfield{journal}{%
  \Doi{10.1103/PhysRevB.93.035304}{\bibinfo {journal} {Phys. Rev. B}}\ }%
  \textbf{\bibinfo {volume} {93}},\ \bibinfo {pages} {035304} (\bibinfo {year}
  {2016})%
  \bibAnnoteFile{NoStop}{Kolasinski2016Lande}%
\bibitem{Zwierzycki2008}%
  \BibitemOpen
  \bibfield{author}{%
  \bibinfo {author} {\bibfnamefont{M.}~\bibnamefont{Zwierzycki}}, \bibinfo
  {author} {\bibfnamefont{P.~A.}\ \bibnamefont{Khomyakov}}, \bibinfo {author}
  {\bibfnamefont{A.~A.}\ \bibnamefont{Starikov}}, \bibinfo {author}
  {\bibfnamefont{K.}~\bibnamefont{Xia}}, \bibinfo {author}
  {\bibfnamefont{M.}~\bibnamefont{Talanana}}, \bibinfo {author}
  {\bibfnamefont{P.~X.}\ \bibnamefont{Xu}}, \bibinfo {author}
  {\bibfnamefont{V.~M.}\ \bibnamefont{Karpan}}, \bibinfo {author}
  {\bibfnamefont{I.}~\bibnamefont{Marushchenko}}, \bibinfo {author}
  {\bibfnamefont{I.}~\bibnamefont{Turek}}, \bibinfo {author}
  {\bibfnamefont{G.~E.~W.}\ \bibnamefont{Bauer}}, \bibinfo {author}
  {\bibfnamefont{G.}~\bibnamefont{Brocks}},\ and\ \bibinfo {author}
  {\bibfnamefont{P.~J.}\ \bibnamefont{Kelly}},\ }%
  \bibfield{journal}{%
  \Doi{10.1002/pssb.200743359}{\bibinfo {journal} {Phys. Stat. Sol.}}\ }%
  \textbf{\bibinfo {volume} {245}},\ \bibinfo {pages} {623} (\bibinfo {year}
  {2008})%
  \bibAnnoteFile{NoStop}{Zwierzycki2008}%
\bibitem{Kolasinski2014Slit}%
  \BibitemOpen
  \bibfield{author}{%
  \bibinfo {author}
  {\bibfnamefont{K.}~\bibnamefont{Kolasi\ifmmode~\acute{n}\else
  \'{n}\fi{}ski}}, \bibinfo {author}
  {\bibfnamefont{B.}~\bibnamefont{Szafran}},\ and\ \bibinfo {author}
  {\bibfnamefont{M.~P.}\ \bibnamefont{Nowak}},\ }%
  \bibfield{journal}{%
  \Doi{10.1103/PhysRevB.90.165303}{\bibinfo {journal} {Phys. Rev. B}}\ }%
  \textbf{\bibinfo {volume} {90}},\ \bibinfo {pages} {165303} (\bibinfo {year}
  {2014})%
  \bibAnnoteFile{NoStop}{Kolasinski2014Slit}%
\bibitem{Kolasinski2015}%
  \BibitemOpen
  \bibfield{author}{%
  \bibinfo {author}
  {\bibfnamefont{K.}~\bibnamefont{Kolasi\ifmmode~\acute{n}\else
  \'{n}\fi{}ski}}\ and\ \bibinfo {author}
  {\bibfnamefont{B.}~\bibnamefont{Szafran}},\ }%
  \bibfield{journal}{%
  \bibinfo {journal} {New J. Phys.}\ }%
  \textbf{\bibinfo {volume} {17}},\ \bibinfo {pages} {063003} (\bibinfo {year}
  {2015})%
  \bibAnnoteFile{NoStop}{Kolasinski2015}%
\bibitem{Jalabert2010}%
  \BibitemOpen
  \bibfield{author}{%
  \bibinfo {author} {\bibfnamefont{R.~A.}\ \bibnamefont{Jalabert}}, \bibinfo
  {author} {\bibfnamefont{W.}~\bibnamefont{Szewc}}, \bibinfo {author}
  {\bibfnamefont{S.}~\bibnamefont{Tomsovic}},\ and\ \bibinfo {author}
  {\bibfnamefont{D.}~\bibnamefont{Weinmann}},\ }%
  \bibfield{journal}{%
  \Doi{10.1103/PhysRevLett.105.166802}{\bibinfo {journal} {Phys. Rev. Lett.}}\
  }%
  \textbf{\bibinfo {volume} {105}},\ \bibinfo {pages} {166802} (\bibinfo {year}
  {2010})%
  \bibAnnoteFile{NoStop}{Jalabert2010}%
\bibitem{Khatua2014}%
  \BibitemOpen
  \bibfield{author}{%
  \bibinfo {author} {\bibfnamefont{P.}~\bibnamefont{Khatua}}, \bibinfo {author}
  {\bibfnamefont{B.}~\bibnamefont{Bansal}},\ and\ \bibinfo {author}
  {\bibfnamefont{D.}~\bibnamefont{Shahar}},\ }%
  \bibfield{journal}{%
  \Doi{10.1103/PhysRevLett.112.010403}{\bibinfo {journal} {Phys. Rev. Lett.}}\
  }%
  \textbf{\bibinfo {volume} {112}},\ \bibinfo {pages} {010403} (\bibinfo {year}
  {2014})%
  \bibAnnoteFile{NoStop}{Khatua2014}%
\bibitem{Bercioux2015}%
  \BibitemOpen
  \bibfield{author}{%
  \bibinfo {author} {\bibfnamefont{D.}~\bibnamefont{Bercioux}}\ and\ \bibinfo
  {author} {\bibfnamefont{P.}~\bibnamefont{Lucignano}},\ }%
  \bibfield{journal}{%
  \bibinfo {journal} {Rep. Prog. Phys.}\ }%
  \textbf{\bibinfo {volume} {78}},\ \bibinfo {pages} {106001} (\bibinfo {year}
  {2015})%
  \bibAnnoteFile{NoStop}{Bercioux2015}%
\end{thebibliography}%

\end{document}